%% file: main.tex
\documentclass[10pt, sigconf,screen]{acmart}

\input{misc/prelude}
\input{misc/notation}

\begin{document}

\title{\snamefull{:} a proof-of-concept for hardware-based virtualization}

\author{Francesco Ciraolo}
\affiliation{\institution{\small Boston University\country{USA}}}
\email{fciraolo@bu.edu}

\author{Mattia Nicolella}
\affiliation{\institution{\small Boston University\country{USA}}}
\email{mnico@bu.edu}

\author{Denis Hoornaert}
\affiliation{\institution{\small TU M\"unchen\country{ Germany}}}
\email{denis.hoornaert@tum.de}

\author{Marco Caccamo}
\affiliation{\institution{\small TU M\"unchen\country{ Germany}}}
\email{marco.caccamo@tum.de}

\author{Renato Mancuso}
\affiliation{\institution{\small Boston University\country{USA}}}
\email{rmancuso@bu.edu}

\input{text/abstract}

\maketitle

\input{text/intro}

\input{text/prerequisites}

\input{text/model}

\input{text/enabled_techniques}

\input{text/conclusion}

\bibliographystyle{plain}
\bibliography{main}

\end{document}

%% file: misc/prelude.tex
\usepackage{lipsum}
\usepackage{xspace}
\usepackage{xcolor}
\usepackage{scalefnt}
\usepackage[super]{nth}
\usepackage{hyperref}
\usepackage{xfp}
\usepackage{booktabs}

\usepackage{tikz}
\usepackage{pgf-umlsd}
\usetikzlibrary{calc}
\usetikzlibrary{positioning}
\usetikzlibrary{ext.paths.ortho} 
\usetikzlibrary {arrows.meta,bending,positioning}
\usetikzlibrary{decorations.pathreplacing,positioning, arrows.meta}

\usepackage[tikz]{bclogo}
\usepackage[framemethod=tikz]{mdframed}
\usepackage[many]{tcolorbox}

\definecolor{bgblue}{RGB}{245,243,253}
\definecolor{ttblue}{RGB}{91,194,224}
\definecolor{matte}{RGB}{247,240,208}
\definecolor{matteDark}{RGB}{227,220,188}
\definecolor{cp_color}{RGB}{0,111,186}

\newtcolorbox{highlightbox}[1][]{
  breakable,
  freelance,
  title=#1,
  colback=matte,
  colbacktitle=matteDark,
  coltitle=black,
  fonttitle=\bfseries,
  bottomrule=0pt,
  boxrule=0pt,
  colframe=white,
  overlay unbroken and first={
    \draw[red!75!black,line width=3pt]
    ([xshift=5pt]frame.north west) --
    (frame.north west) --
    (frame.south west);
    \draw[red!75!black,line width=3pt]
    ([xshift=-5pt]frame.north east) --
    (frame.north east) --
    (frame.south east);
  },
  overlay unbroken app={
    \draw[red!75!black,line width=3pt,line cap=rect]
    (frame.south west) --
    ([xshift=5pt]frame.south west);
    \draw[red!75!black,line width=3pt,line cap=rect]
    (frame.south east) --
    ([xshift=-5pt]frame.south east);
  },
  overlay middle and last={
    \draw[red!75!black,line width=3pt]
    (frame.north west) --
    (frame.south west);
    \draw[red!75!black,line width=3pt]
    (frame.north east) --
    (frame.south east);
  },
  overlay last app={
    \draw[red!75!black,line width=3pt,line cap=rect]
    (frame.south west) --
    ([xshift=5pt]frame.south west);
    \draw[red!75!black,line width=3pt,line cap=rect]
    (frame.south east) --
    ([xshift=-5pt]frame.south east);
  },
}

\usepackage{pifont}

\graphicspath{{imgs/}}
\newcommand{\fig}[1]{Fig.~\ref{fig:#1}}

\newcommand{\eg}{{\it e.g.,}\xspace}

\newcommand{\ie}{{\it i.e.,}\xspace}
\newcommand{\Ie}{{\it I.e.,}\xspace}

\newcommand{\ci}{{\it (i) }}
\newcommand{\cii}{{\it (ii) }}
\newcommand{\ciii}{{\it (iii) }}
\newcommand{\civ}{{\it (iv) }}
\newcommand{\cv}{{\it (v) }}

\definecolor{colori}{HTML}{2EEA9D}
\definecolor{colorii}{HTML}{1E88E5}
\definecolor{coloriii}{HTML}{FFC107}
\definecolor{coloriv}{HTML}{D81531}
\definecolor{colorv}{HTML}{DC1697}
\definecolor{colorvi}{HTML}{004D40}

\usepackage{subcaption}
\usepackage[]{todonotes}

\usepackage{float}

%% file: misc/notation.tex
\newcommand{\para}[1]{\smallskip\noindent\textbf{#1}.}

\newcommand{\step}[1]{\textbf{(#1)}\xspace}

\newcommand{\wrt}{w.r.t.\xspace}

\newcommand{\sect}{Sec.}

\newcommand{\va}{VA\xspace}
\newcommand{\valong}{Virtual Address\xspace}
\newcommand{\pa}{PA\xspace}
\newcommand{\palong}{Physical Address\xspace}
\newcommand{\ipa}{IPA\xspace}
\newcommand{\ipalong}{Intermediate Physical Address\xspace}
\newcommand{\pfn}{PFN\xspace}
\newcommand{\pfnlong}{Page Frame Number\xspace}
\newcommand{\mmu}{MMU\xspace}
\newcommand{\mmulong}{Memory Management Unit\xspace}
\newcommand{\tlb}{TLB\xspace}
\newcommand{\tlblong}{Translation Look-aside Buffer\xspace}
\newcommand{\pgd}{PGD\xspace}

\newcommand{\vm}{VM\xspace}
\newcommand{\vmlong}{Virtual Machine\xspace}

\newcommand{\soc}{SoC\xspace}

\newcommand{\axi}{AXI\xspace}

\newcommand{\ace}{ACE\xspace}
\newcommand{\acelong}{\axi Coherency Extension\xspace}

\newcommand{\pe}{PE\xspace}

\newcommand{\cpu}{CPU\xspace}

\newcommand{\gpu}{GPU\xspace}

\newcommand{\fpga}{FPGA\xspace}

\newcommand{\arm}{AArch\xspace}

\newcommand{\zculong}{Zynq UltraScale+ MPSoC ZCU102\xspace}
\newcommand{\os}{OS\xspace}
\newcommand{\oslong}{Operating System\xspace}
\newcommand{\pte}{PTE\xspace}
\newcommand{\ptelong}{Page Table Entry\xspace}
\newcommand{\io}{I/O\xspace}

\newcommand{\kb}{~KB\xspace}
\newcommand{\mb}{~MB\xspace}

\newcommand{\snamefull}{Light Virtualization\xspace}
\newcommand{\sname}{LightV\xspace}

\newcommand{\master}{master}

%% file: text/abstract.tex
\begin{abstract}
	Virtualization has become widespread across all computing environments, from
	edge devices to cloud systems. Its main advantages are resource management
	through abstraction and improved isolation of platform resources and
	processes. However, there are still some important tradeoffs as it requires
	significant support from the existing hardware infrastructure and negatively
	impacts performance. Additionally, the current approaches to resource
	virtualization are inflexible, using a model that doesn't allow for dynamic
	adjustments during operation. This research introduces \snamefull (\sname), a
	new virtualization method for commercial platforms. \sname uses programmable
	hardware to direct cache coherence traffic, enabling precise and seamless
	control over which resources are virtualized. The paper explains the core
	principles of \sname, explores its capabilities, and shares initial findings
	from a basic proof-of-concept module tested on commercial hardware.
\end{abstract}

%% file: text/intro.tex
\section{Introduction}
    \label{sec:introduction}

    Virtualization is a staple of modern computing systems.
    In the last decade, support and use cases for hardware-aided virtualization have substantially proliferated, touching a wide range of real-world deployment from large-scale enterprise and cloud infrastructure~\cite{virt_cloud}, to general-purpose consumer computing~\cite{wolf2006virtualization}, to embedded systems~\cite{virt_embedded,shedding-light}. Indeed, virtualization support offers flexible hardware provisioning, 
    allows abstracting away resources, managing their visibility to guest \os's, and isolating them as needed to achieve multi-tenant/multi-domain setups with spatial and performance isolation properties. 
        
    Unfortunately, the virtualization superpower comes with a hefty price.
    First, the need to introduce one (or more) layers of software and hardware often comes with noticeable overheads~\cite{virt_overhead}.
    Moreover, the additional software layers add complexity and widen the security attack surface~\cite{security_hyper}.
    Second, current support for virtualization is \textit{"binary"}, \ie either everything or nothing in a system is virtualized.
    Although some hypervisors can be dynamically activated after boot~\cite{jailhouse}, once a hypervisor is active, there is little to no control over what resources are virtualized and which ones are not.
    For instance, any memory page accessed by a guest OS must be mapped to physical memory by the hypervisor. 
    
    The aforementioned shortcomings of traditional memory and \io virtualization technology arise from its reliance on page table-based address translation.
    However, what if one could \emph{steer} address translation using arbitrary logic?
    In this case, intermediate-to-physical address resolution could be dynamically defined on the fly.
    
    In this paper, we demonstrate that this is already possible in commercial hardware by surgically interacting with the cache coherence traffic originating from a \cpu cluster.
    We refer to one such approach to enact resource virtualization as \emph{\snamefull}, or \sname for short.
    \sname enables us to re-think virtualization support and opens the doors to untapped resource management paradigms.
    To this end, we present the proof-of-concept implementation of a hardware module capable of providing basic \sname primitives and the preliminary results we obtained on an initial single-page virtualization use case.

%% file: text/prerequisites.tex
\section{Ancillary Concepts}

    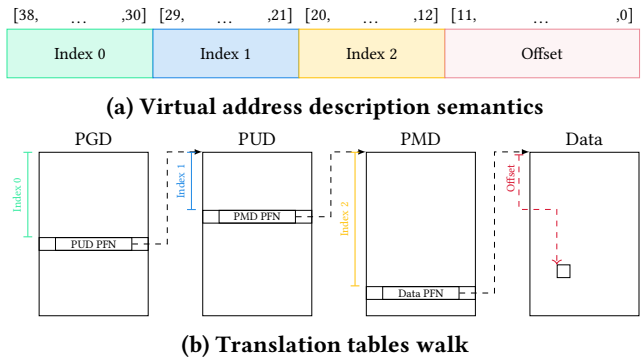
\begin{figure}
        \begin{subfigure}{\linewidth}
        \centering
            \resizebox{\linewidth}{!}{\input{imgs/va}}
            \caption{Virtual address description semantics}
            \label{fig:va}
            \Description{Virtual address description semantics}
        \end{subfigure}
        \begin{subfigure}{\linewidth}
        \centering
            \resizebox{\linewidth}{!}{\input{imgs/translation}}
            \caption{Translation tables walk}
            \label{fig:translation}
            \Description{Translation tables walk}
        \end{subfigure}
        \caption{\va to \pa translation in \arm{64}, 4\kb page granularity and \nth{4} level of translation folded.}
        \label{fig:mmu}
        \vspace{-0.5cm}
    \end{figure}

    This section covers fundamental concepts that are key to understanding \sname.
    The terminology used in this section is based on \arm{64} systems to ease the transition between the description of the concepts and our prototype.
    
    \para{Address Translation}
        Virtual memory is ubiquitous in modern general-purpose \oslong (\os).
        Its objective is to map a contiguous set of \valong{es} (\va) seen by a process to a set of \palong{es} (\pa) that may not be contiguous at a given granule (page) size.

        To allow for the implementation of arbitrary mappings, virtualization is achieved via page tables, i.e., a $l$-level $n$-ary radix tree walked by segmenting the \va in several indices. \fig{mmu} illustrates how a page table walk is performed on \arm{64} systems. A page table is looked up using said indices, shown in green, blue, or orange in \fig{mmu}. At each translation step, the content of the \ptelong (\pte) is a \emph{pointer} to the following translation step or a leaf page. Specifically, the \pte contains the \pfnlong (\pfn) of the next page table or of the final mapped data page.
        Page table walks are handled in hardware by the \mmulong (\mmu) upon each \cpu-originated \va memory access. As translations may occur frequently and are time costly, \mmu{s} are often equipped with multiple levels of \tlblong (\tlb) to cache translation results. Usually, the \mmu{s} can access the same cache hierarchy as the CPU load/store unit.
    
    \para{Type 1 Hypervisors}
        The terminology \emph{``Type 1 Hypervisors''} refers to a class of bare-metal hypervisors that directly interfaces a hosted \vmlong (\vm) with the hardware layer. They can partition system resources (memory, I/O devices) to the \vm{s} by leveraging specialized hardware support that introduces yet another level of indirection between \va{s} as translated by \vm{s} and real \pa{s}. Similarly to the \va-to-\pa translation described above, a secondary translation stage is introduced, leveraging page tables again. In this case, \os{es} in \vm{s} are said to translate \va{s} to \ipalong{s} (\ipa{s}) whereas, hypervisor-side page tables translate \ipa{s} to \pa{s}.        
        Of course, this extra translation layer introduces additional overheads, as up to two full page table walks may be performed for a single \va memory access.
        Moreover, hypervisor bookkeeping operations and the handling of \vm exits steal \cpu time from the hosted \vm{s}.
        
    \para{Cache Coherence}
        In modern multi-processors \soc{s} where the cache hierarchy is a key performance enabler, ensuring a consistent view of the memory to all attached \pe{s} is crucial.
        Snoop-based coherence is a common coherence paradigm.
        The core idea is to attach a few bits to each cache line to encode its state (\eg \emph{Modified}, \emph{Shared}, \emph{Invalid}) in the cache hierarchy.
        The state of each cache line in the hierarchy is continuously updated as they are accessed by the \pe{s}.

        Under snoop-based coherence, \pe-originated read/write requests prompt a coherence controller (\eg interconnect) to broadcast commands to all other attached \pe{s}.
        These commands directly affect the state of the targeted cache lines in the \pe{s}' caches.
        For instance, a read-miss event from \pe{1} will prompt the controller to broadcast a snoop request.
        If the snoop results in a hit in \pe{2}'s cache, the line state in \pe{2}'s cache will be altered---\eg \ci from clean and exclusive to clean and shared or \cii from modified and exclusive to invalid~\cite{MESI_protocol}.
        Conversely, in case of a miss in the cache of all the other coherent PEs, the request is forwarded to the next level of the cache hierarchy (\eg another cache level or main memory), and the line is allocated in \pe{1}'{s} cache.
        
        On \arm{64} \soc{s}, the \acelong (\ace) is an established bus-level interface to implement snoop-based coherence protocols. For instance, ACE-enabled \cpu-side cache controllers and interconnects (\eg ARM CCI-400~\cite{CCI-400}) are the basic components of a coherent multi-cluster big.LITTLE architecture~\cite{ARM-big-LITTLE, chung2012heterogeneous}.

    \para{Coherence Backstabbing}
        Simply put, the idea behind \emph{Coherence Backstabbing} as originally proposed in~\cite{CAESAR} is to hijack the coherence protocol for purposes other than maintaining coherence among a multi-\pe system where one of the \pe{s} is a coherent \fpga~\cite{enzian2020cidr}.
        Coherence backstabbing leverages the idea that a coherent \pe can \emph{pretend} to have ownership of a cache line, fabricate its content, and feed it to the other \pe.
        In other words, by \emph{``placing itself on the back of the coherence protocol''}~\cite{CAESAR}, a custom user-defined hardware module can seamlessly intercept, modify, and reroute \emph{any} coherent traffic---\eg cache line refills.

%% file: imgs/va.tex
\begin{tikzpicture}

\node[rectangle,
	draw = colori,
	text = black,
	fill = colori!20!white,
    anchor=south west, 
	minimum width = 3cm,
	minimum height = 1cm] (index0) at (0,0) {Index 0};
\node[rectangle,
	draw = colorii,
	text = black,
	fill = colorii!20!white,
    anchor=south west, 
	minimum width = 3cm,
	minimum height = 1cm] (index1) at (3,0) {Index 1};
\node[rectangle,
	draw = coloriii,
	text = black,
	fill = coloriii!20!white,
    anchor=south west, 
	minimum width = 3cm,
	minimum height = 1cm] (index2) at (6,0) {Index 2};
\node[rectangle,
	draw = coloriv!50!white,
	text = black,
	fill = coloriv!5!white,
    anchor=south west, 
	minimum width = 4cm,
	minimum height = 1cm] (offset) at (9,0) {Offset };

\node [anchor=south west] () at (index0.north west) {[38,};
\node [anchor=south] () at (index0.north) {\dots};
\node [anchor=south east] () at (index0.north east) {,30]};

\node [anchor=south west] () at (index1.north west) {[29,};
\node [anchor=south] () at (index1.north) {\dots};
\node [anchor=south east] () at (index1.north east) {,21]};

\node [anchor=south west] () at (index2.north west) {[20,};
\node [anchor=south] () at (index2.north) {\dots};
\node [anchor=south east] () at (index2.north east) {,12]};

\node [anchor=south west] () at (offset.north west) {[11,};
\node [anchor=south] () at (offset.north) {\dots};
\node [anchor=south east] () at (offset.north east) {,0]};
 
\end{tikzpicture}

%% file: imgs/translation.tex
\begin{tikzpicture}

\node[rectangle,
	draw = black,
	text = black,
    anchor=south west, 
	minimum width = 2cm,
	minimum height = 3cm] (pgd) at (0,0) {};
\node [anchor=south] () at (pgd.north) {PGD};
 
\node[rectangle,
	draw = black,
	text = black,
    anchor=south west, 
	minimum width = 2cm,
	minimum height = .1cm] (pte0) at (0,1.2) {};

\node[rectangle,
	draw = black,
	text = black,
    font=\tiny, 
	minimum width = 1.4cm,
	minimum height = .1cm] (pfn0) at (pte0.center) {};

\node[rectangle,
	text = black,
    font=\tiny, 
	minimum width = 1.6cm,
	minimum height = .1cm] () at (pte0.center) {PUD PFN};

\draw[draw=colori, |-|] ($(pte0.north west)!.2cm!90:(pgd.north west)$) --node[font=\tiny, above, sloped, swap, text=colori]{Index 0} ($(pgd.north west)!.2cm!-90:(pte0.north west)$);

\node[rectangle,
	draw = black,
	text = black,
    anchor=south west, 
	minimum width = 2cm,
	minimum height = 3cm] (pud) at (3,0) {};
\node [anchor=south] () at (pud.north) {PUD};

\draw [dashed, -latex] (pfn0.east) -|- (pud.north west);

\node[rectangle,
	draw = black,
	text = black,
    anchor=south west, 
	minimum width = 2cm,
	minimum height = .1cm] (pte1) at (3,1.7) {};

\node[rectangle,
	draw = black,
	text = black,
    font=\tiny, 
	minimum width = 1.4cm,
	minimum height = .1cm] (pfn1) at (pte1.center) {};

\node[rectangle,
	text = black,
    font=\tiny, 
	minimum width = 1.6cm,
	minimum height = .1cm] () at (pte1.center) {PMD PFN};

\draw[draw=colorii, |-|] ($(pte1.north west)!.2cm!90:(pud.north west)$) --node[font=\tiny, above, sloped, swap, text=colorii]{Index 1} ($(pud.north west)!.2cm!-90:(pte1.north west)$);


\node[rectangle,
	draw = black,
	text = black,
    anchor=south west, 
	minimum width = 2cm,
	minimum height = 3cm] (pmd) at (6,0) {};
\node [anchor=south] () at (pmd.north) {PMD};

\draw [dashed, -latex] (pfn1.east) -|- (pmd.north west);

\node[rectangle,
	draw = black,
	text = black,
    anchor=south west, 
	minimum width = 2cm,
	minimum height = .1cm] (pte2) at (6,0.3) {};

\node[rectangle,
	draw = black,
	text = black,
    font=\tiny, 
	minimum width = 1.4cm,
	minimum height = .1cm] (pfn2) at (pte2.center) {};

\node[rectangle,
	text = black,
    font=\tiny, 
	minimum width = 1.6cm,
	minimum height = .1cm] () at (pte2.center) {Data PFN};

\draw[draw=coloriii, |-|] ($(pte2.north west)!.2cm!90:(pmd.north west)$) --node[font=\tiny, above, sloped, swap, text=coloriii]{Index 2} ($(pmd.north west)!.2cm!-90:(pte2.north west)$);


\node[rectangle,
	draw = black,
	text = black,
    anchor=south west, 
	minimum width = 2cm,
	minimum height = 3cm] (data) at (9,0) {};
\node [anchor=south] () at (data.north) {Data};

\draw [dashed, -latex] (pfn2.east) -|- (data.north west);

\node[rectangle,
	draw = black,
	text = black,
    anchor=south west, 
	minimum width = .1cm,
	minimum height = .1cm] (datab) at (9.5,0.7) {};

\draw [draw=coloriv, dashed, -latex, |-> ] ($(data.north west)!.2cm!-90:(datab.north west)$) |-| (datab.north west);
\node [font=\tiny,
    text=coloriv,
    rotate=90,
    anchor=south east
] () [above left =  -0.1 cm and 0.15 cm  of data.north west] {Offset};

\end{tikzpicture}

%% file: text/model.tex
\section{Switching to a lighter weapon}
    \label{sec:model}

    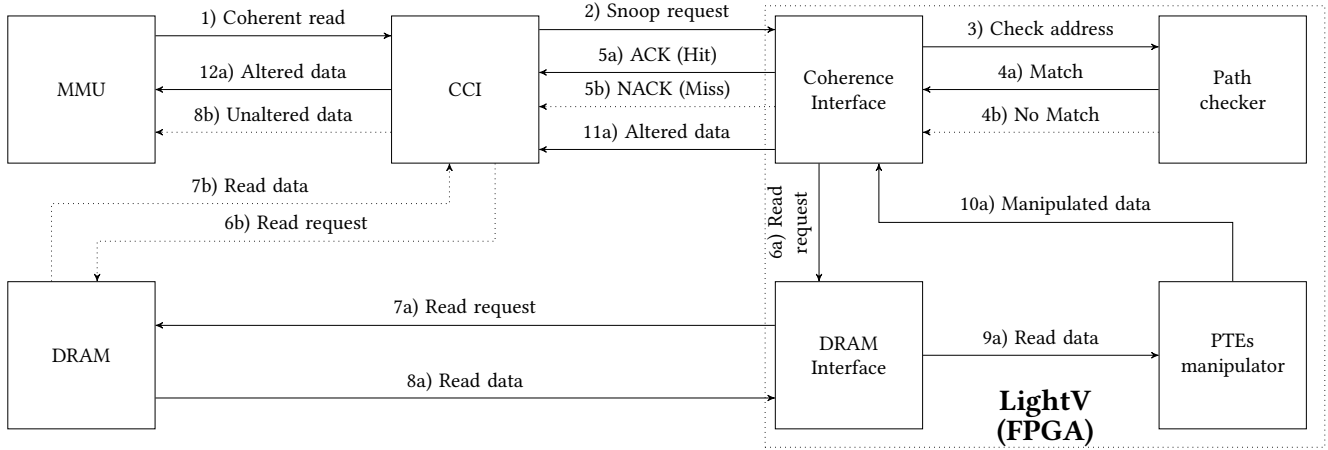
\begin{figure*}
      \centering
      \resizebox{\linewidth}{!}{\input{imgs/svn_skeleton}}
      \vspace{-1cm}
      \caption{Skeleton and fundamental operations of a basic \sname module}
      \label{fig:skeleton}
      \vspace{-0.2cm}
    \end{figure*}
    
    As an alternative approach to traditional virtualization, we propose \snamefull (\sname). \sname is a paradigm to implement hardware modules that build atop the concept of coherence backstabbing~\cite{CAESAR} to arbitrarily manipulate \va-to-\pa translation in a manner that remains seamless to software components.
    \sname can achieve \emph{transparent}, \emph{temporary}, and \emph{on-demand} control over the translation process. This is achieved through observation and selective manipulation of the data payloads returned via coherence in response to \mmu-generated snoops during the page table walk.
    This level of control allows for a \emph{``true''} in-hardware virtualization with little support required from the vendors.\footnote{Although, the model can be further simplified and improved if additional support is provided in hardware, as discussed in \sect~\ref{subsec:on_the_vendors}.}

    \subsection{System Model}
        \label{subsec:system_model}

        A few features are required for \sname to be deployed on a \soc.
        The system must include one or more \pe{s} (\eg \cpu, \cpu{s} cluster, \gpu) supporting memory virtualization (\ie equipped with an \mmu and \tlb{s}).
        The \pe-side cache must be capable of handling coherence and be connected to a coherent controller (\eg coherent interconnect) employing a snoop-based protocol. The latter interfaces the coherence domain with main memory.
        \mmu-originated accesses to \pte{s} during page table walks are treated as cacheable, and therefore generate snoops that are broacasted to all other coherent \pe{s} upon a \pe-side cache miss.
  
        We make no assumptions on whether the module is implemented as soft-logic in a programmable logic fabric (as in \cite{CAESAR}) or as a fixed dedicated piece of hardware. To illustrate the basic principle, however, we hereby make the following simplifying assumptions. We discuss in \sect~\ref{subsec:challenges} how these can be substantially relaxed.
        \begin{enumerate}
        
            \item No \pte{s} are cached. All \mmu-originated accesses miss in the \pe-side cache and the corresponding snoops are broadcasted to the other coherent \pe{s}.
            \item The target \va range is \emph{isolated} and \emph{pre-mapped}. \Ie no other valid \va{s} share the same first index, and all the intermediate \pte{s} have been pre-populated before \sname is activated.
            \item The \pgd of the target address space that contains the target \va{s} is known before \sname is activated.
            \item The new destination \pa{s} and desired mapping attributes are provided by the end-user. 
        \end{enumerate}

    \subsection{\sname Operation}\label{sec:snv_op}

        As shown in \fig{skeleton}, the \sname module activates each time it receives a snoop command.
        This command is analyzed and compared (step \step{3}) to determine if the snoop command results from a \pte access that was missed in the cache.
        At the very beginning, \ie right after the \sname module has been activated, the check is performed against the known \pgd address plus possible offsets. Later, the check will also include any intermediate PFNs of valid page tables obtained in the previous translation steps for the \va{s} of interest.
        
        If the snooped address indicates access to a \pte of interest, the \sname module acknowledges the request (step \step{5a}), indicating to the cache coherent interconnect (CCI) that it has the most up-to-date data payload.
        From then on, the latter waits for the data to be served at step \step{11a}.
        Meanwhile, the \sname module fetches from main memory the payload of the requested \pte (steps \step{7a} and \step{7b}).
        The payload is passed (step \step{9a}) to the \pte{s} manipulator. Here, it can be altered as necessary following arbitrary logic before being fed to the coherent controller (step \step{11a}) and eventually the \mmu (step \step{12a}).
        Note that, in parallel to this process, the \sname module internally maintains a \emph{context cache} within the translation path checker. This is useful for storing additional metadata about the current translation.
        
        Conversely, if the snoop is not relevant for the translation of any of the \va{s} of interest (step \step{4b}), the \sname module provides a negative acknowledgment to the CCI (step \step{5b}).
        From there, a "text-book" access to the page tables is performed by fetching the non-altered tables directly from main memory (steps \step{6b} and \step{7b}).

        Note that, with no \pte{s} cached on the \pe side, the routine will be repeated once per active page table level to complete a full-page table walk---\eg up to four times in \arm{64}.

    \subsection{Relaxing \sname Assumptions} 
        \label{subsec:challenges}

        The assumptions listed in \sect~\ref{subsec:system_model} are representative of the various technical challenges that remain to be addressed.
        This section explains the rationale behind each assumption and how they can be enforced or relaxed.

        \para{Assumption (1)} 
            Disabling the caching of \pte{s} in the cache hierarchy is undesirable, as it is crucial to hide costly translation latencies.
            On the other hand, if intermediate \pte{s} passed to the \pe might be cached, then not all the \mmu-generated \pte accesses will result in \pe-side cache misses and snoops.             This \emph{partial observability} means that \sname is oblivious to the walks' progress and, thus, cannot determine how to appropriately alter the translations when the next relevant \pte miss occurs. 
            
            Fortunately, the \pte addresses returned by the \sname module do not have to be valid \pa{s}---because accesses to them will be intercepted when later snooped. Thus, we can inject a \emph{watermark} in the payload of all the manipulated \pte{s} served to the \pe.
            In addition to encoding the translation step number, the watermark can be looked up in the internal context cache (\sect~\ref{sec:snv_op}) to recover several other attributes necessary for proper \sname operation.

        \para{Assumption (2)}
            The reason for \textbf{Assumption (2)} is to avoid that also neighbor \va{s} sharing intermediate page tables could trigger the same issue raised if Assumption (3) was lifted. Thus, it can be relaxed if Assumption (3) is lifted. 

        \para{Assumption (3)}
            Assuming pre-populated \va mappings before \sname is activated is a simplifying assumption because Linux employs demand paging.
            In brief, if the \sname module serves a manipulated \pte but no subsequent page is allocated (present bit is cleared), the \mmu generates a fault. When handling the fault, the kernel retrieves the potentially invalid \pa of the \pte and might attempt to populate it.
            However, as the address is invalid and only makes sense when looked up in the context cache, it might cause a kernel-side page fault. This problem can be solved with an \sname module capable of consistently manipulating both user-space and kernel-space \va-to-\pa translations.

    \subsection{Enabling Virtualization at All Levels}

        The proposed approach does not preclude the system from running a hypervisor.
        In fact, the mechanism presented earlier can be extended to hypervisor translations, too.
        Instead of intervening only during the \va-to-\pa translation, \sname could also monitor and alter \ipa-to-\pa translations.

%% file: imgs/svn_skeleton.tex
\newcommand{\nodedistance}{4 cm}
\newcommand{\squareside}{2.5 cm}
\tikzstyle{arrow style}=[text width=(\nodedistance - .75 cm),midway,above, minimum height=0cm, minimum width=0cm]
\begin{tikzpicture}[anchor=center, minimum width=\squareside, text width=(\squareside - .75 cm), align=center, minimum height=\squareside,auto,>=stealth']

    \node(rect) [draw] (MMU) {MMU};
    \node(rect) [draw, right = \nodedistance of MMU] (CCI) {CCI};
    \node(rect) [draw, below = \nodedistance/2 of MMU] (DRAM) {DRAM};

    \node(rect) [draw, right = \nodedistance of CCI] (ace) {Coherence Interface};

    \node(rect) [draw, below = \nodedistance/2 of ace] (axi) {DRAM Interface};

    \node(rect) [draw, right = \nodedistance of ace] (pathcheck) {Path checker};
    \node(rect) [draw, right = \nodedistance of axi] (manipulator) {PTEs\\manipulator};
    


    \newcommand\svncol{2}
    \newcommand\svnrow{2}

    \newcommand\svnwidth{\nodedistance * (\svncol - 1) + \squareside * \svncol + .5 cm}
    \newcommand\svnheight{\nodedistance/2 * (\svnrow - 1) + \squareside * \svnrow + .5 cm}

    \newcommand{\svnside}{\nodedistance * 1 + \squareside * 2 + .5 cm}
    \newcommand{\svndelta}{\squareside + 1.3 cm}

    \node(rect) [draw, dotted, below right = - (\svndelta) of ace, minimum width=\svnwidth, minimum height=\svnheight] (pd) {};
    \node [above= -.75 cm of pd.south] () {\textbf{\LARGE \sname (FPGA)}};

    \draw[->] (MMU.36) -- node [arrow style] {1) Coherent read} (CCI.144) {};
    \draw[->] (CCI.39) -- node [arrow style] {2) Snoop request} (ace.141) {};
    \draw[->] (ace.30) -- node [arrow style] {3) Check address} (pathcheck.150) {};
    
    \draw[->] (pathcheck) -- node [arrow style] {4a) Match} (ace) {};
    \draw[->] (ace.167) -- node [arrow style] {5a) ACK (Hit)} (CCI.13) {};
    \draw[->] (ace.248) -- node [arrow style, rotate=90, text width=1.5cm, align=center] {6a) Read request} (axi.112) {};
    \draw[->] (axi.158) -- node [arrow style] {7a) Read request} (DRAM.22) {};
    \draw[->] (DRAM.330) -- node [arrow style] {8a) Read data} (axi.210) {};
    \draw[->] (axi) -- node [arrow style] {9a) Read data} (manipulator) {};
    \draw[->] (manipulator) |- node [arrow style] {10a) Manipulated data}| (ace.292) {};
    \draw[->] (ace.219) -- node [arrow style] {11a) Altered data} (CCI.321) {};
    \draw[->] (CCI) -- node [arrow style] {12a) Altered data} (MMU) {};

    
    \draw[->, dotted] (pathcheck.210) -- node [arrow style] {4b) No Match} (ace.330) {};
    \draw[->, dotted] (ace.193) -- node [arrow style] {5b) NACK (Miss)} (CCI.347) {};
    \draw[->, dotted] (CCI.292) |- node [arrow style] {6b) Read request}|[ratio=0.66] (DRAM.78);
    \draw[->, dotted] (DRAM.112) |- node [arrow style] {7b) Read data}|[ratio=.66] (CCI.258);
    \draw[->, dotted] (CCI.210) -- node [arrow style] {8b) Unaltered data} (MMU.330) {};

\end{tikzpicture}

%% file: text/enabled_techniques.tex
\section{On the Battlefield}
    We have implemented a first prototype of \sname as outlined in \sect~\ref{sec:model} on the \zculong development board.
    This board features a \soc composed of four ARM64 Cortex-A53 \cpu cores and a coherent \fpga which hosts our \sname prototype.
    This initial prototype implements the watermarking feature discussed in \sect~\ref{subsec:challenges} to address Assumption~(1).
    However, the prototype still requires an isolated target VA, as per Assumption~(2) and that the \va mappings are pre-populated---Assumption~(3).
    This prototype focuses on virtualizing a subset of pages in a Linux user-space process. This section presents early results \wrt measured overhead introduced by \sname.

    \subsection{Overhead}

        \begin{table}[ht]
            \centering
            \begin{tabular}{cccc}
                Linux & Passive SNV & Active SNV & Jailhouse  \\
                 \toprule
                 $2.39 \pm 0.035s$ & $2.4 \pm 0.052s$ & $2.38 \pm 0.04s$ & $2.38 \pm 0.034s$ \\
                 \bottomrule
            \end{tabular}
            \caption{Avg/min/max runtime of the histogram benchmark. "Active \sname{"} has single-page virtualization.}
            \label{tab:OH-SNV}
            \vspace{-0.8cm}
        \end{table}

        To measure the overheads of single-page virtualization, the RGB histogram benchmark was chosen, with an $\sim$55.6\mb input image. This application is characterized by a large pool of cold pages for the image data, warmer pages for the code itself, and a single hot page for the aggregated data of the histogram; the \va of the latter is the target for \sname.

        The benchmark is executed under three scenarios, shown in Table~\ref{tab:OH-SNV}:
        \textbf{Linux}: Linux without \sname being implemented in the \fpga, averaging  $2.39 \pm 0.035s$; 
        \textbf{Passive \sname}: our \sname module is instantiated in the \fpga as a coherent {\master} but it has no target \va{s}, averaging $2.38 \pm 0.04s$;
        \textbf{Active \sname}: \sname implemented and enabled to virtualize a single page, averaging $2.38 \pm 0.04s$.
        Comparing the \textbf{Linux} and \textbf{Passive \sname} cases captures the overhead introduced by adding a {\master} to the coherence domain.
        As reported in Table~\ref{tab:OH-SNV}, the overhead is negligible.
        Indeed, \textbf{Active \sname} introduces overheads that are in line with the 1\% figure reported by previous works~\cite{shedding-light} for common type-1 hypervisors---\eg Jailhouse, which we also evaluate in Table~\ref{tab:OH-SNV}.
    
    \subsection{Ascertained Novel Features}
    
        \para{Selective Virtualization}
            Our technology allows virtualizing just what is needed with the granularity of a single process/OS page. Unlike traditional virtualization, any translation outside the range of interest will be minimally impacted since no additional steps are added in this case---step~\step{5b} in \fig{skeleton}.
            Moreover, since the \sname machinery is synthesized in FPGA, no virtualization management cost, in the form of extra instructions to be executed on the CPU, needs to be paid.
            Finally, no additional main memory will be allocated for 1:1 flat translations.
        
        \para{Dynamic Translation Semantics}
            As noted previously, the content of traditional translation tables can be changed at any time, allowing more sophisticated mapping.
            Usually, however, this is an expensive operation, and use cases beyond allocation/deallocation of new areas are seldom performed.
            Conversely, under \sname, the rule-based, dynamic translation is more lightweight, and arbitrarily complex semantics can be implemented in the \fpga.

%% file: text/conclusion.tex
\section{Next Steps}
    While the proof-of-concept prototype presented in this paper is a promising stepping stone toward full-fledged \sname, much more research is needed to fully explore the many new avenues opened by this technique.
    
    On the one hand, our immediate future work is focused on tackling the key challenges discussed in \sect~\ref{subsec:challenges}. To this end, we aim to design a \sname module capable of providing virtualization for an arbitrarily large portion of the virtual addressing space of a user-space application running in a typical OS, \eg Linux. On the other hand, we envision that several potential paradigms are within reach after that, as summarized below.

    \subsection{Future Paradigms Enabled by \sname}
    
        \para{Seamless pages migration}
            Migrating one or more pages of a user process or guest OS is not uncommon and done for multiple reasons---\eg moving a page closer to the physical core in a NUMA configuration, exploiting faster/non-volatile memory. Nonetheless, page migration is an expensive operation, requiring the processes involved to be blocked while busy-copy or DMA operations are performed and the affected page table(s) are updated.
            
            With \sname, it is possible to seamlessly migrate pages. Upon migration, \ci a DMA operation is issued to copy a given page; simultaneously, \cii any \tlb entry for the page is invalidated by the \sname module, which \ciii manipulates future translations to target the new page location. Before the DMA operation completes, \civ any access performed by the \textbf{running} processes is captured at the coherence level and replied with the data of the source page. \cv Once the DMA completes, the old page can be reclaimed, and accesses to the new page are not captured anymore.
        
        \para{Multiple Hypervisors Coexistence}
            State-of-the-art virtualization support allows for, at most, a single type-1 hypervisor active in the system. By leveraging \sname, it is also possible to manipulate hypervisor-side translations. Thus, it is foreseeable that multiple type-1 hypervisors could be allowed to operate simultaneously on the same physical platform.
        
        \para{\io Device Sharing}
            Type-1 hypervisors can statically assign a given \io resource in pass-through mode to a single \vm. Exporting the same device to multiple \vm{s} requires device emulation or advanced support like SR-IOV~\cite{sr_iov}.
            
            We envision that \sname could enable \io device virtualization in two ways. \ci First, one could replace the physical \pfn of the \io device's control registers and buffers with dedicated addresses in the \fpga aperture while having a hardware module in charge of selectively forwarding commands and data to the true aperture of the memory-mapped \io device.
            \cii Second, one could change the attribute set of the mapped physical device aperture to mark any page within that as cacheable. This way, accessing them will cause snoops that can be intercepted and manipulated. In this case, cache maintenance operations are also issues on the coherence interface to prevent stale data from being retained in any \cpu-side caches.
            
            The main advantage of scenario \ci is that it provides deterministic and direct knowledge of a \vm{s'} access to the device's registers. Indeed, the intercepted read/write bus transaction contains the exact target register address and the same width. Conversely, in scenario \cii, a cache line snoop can only target a cache line size-aligned address. Also, recovering the type of access (read vs. write) could prove challenging.       
            On the other hand, approach \cii allows a more compact design and seamless interaction without dedicating a portion of the \fpga address range to the virtualized device.
        
        \para{Code Execution and Interrupt Virtualization}
            For \sname to ultimately offer full-fledged virtualization support, the ability to execute management code on the \cpu is also desirable, as typical in type-1 hypervisors. It might appear that \sname only allows some address translation \emph{tricks}. Nonetheless, the translation tricks can be leveraged to unlock code execution in \sname. This is because, by carefully steering the physical translation of instruction fetches, \sname can direct the execution flow of the \cpu to any code in memory.
            
            But that's not all. A \sname module could simply replace, at the coherence level, the payload of the next cache line of instructions that is about to be executed by the \cpu, by hijacking the corresponding snoop request. The instructions served instead might be fetched from memory or procedurally generated on the fly as needed. Remarkably, by applying this approach when interrupt handling routines are executed, interrupt virtualization in \sname can be achieved as well.
            
    \subsection{On the vendors}
        \label{subsec:on_the_vendors}
        
        While our solution is thought to be seamless and transparent to be integrates within existing off-the-shelf systems, there is some hardware support that could be easily implemented by the vendors and could make the development easier and the tool more powerful.
        Some of the most relevant challenges depend on \textit{guessing} the intention of the main CPU cluster --- core or MMU --- with a given snoop request; in fact, the latter misses information about the exact address requested and even the nature --- read or write --- of the access.
        
        Especially in the optic of the close approach of the Compute Express Link standard, it could be valuable to have some additions to the coherence protocol:
        the snoop requests should have information related to the \ci memory access direction, \cii the offset --- in the cache line --- of the requested data and \ciii the source of the request --- \ie the core or MMU ID --- in a standard way; also, as an optional feature, \civ even in the case of a cache hit, at any level, a snoop should be performed and potentially waited for, so that the coherent hardware hypervisor can apply a veto and replace the content of the cache line.